\documentclass[%
 reprint,
 amsmath,amssymb,
 pra,
 showpacs
]{revtex4-1}

\usepackage{graphicx}
\usepackage{dcolumn}
\usepackage{bm}
\usepackage{hyperref}

\usepackage{blindtext}
\usepackage{braket}
\usepackage{booktabs}
\usepackage{SIunits}

\usepackage{color}
\usepackage{times}

\newcommand{\ii}{\mathrm{i}}
\newcommand{\ee}{\mathrm{e}}

\renewcommand{\Re}{\operatorname{Re}}

\newcommand{\Tr}{\operatorname{Tr}}
\newcommand{\Ven}{\hat V_{\mathrm{en}}}
\newcommand{\he}{\hat h_{\mathrm{e}}}
\newcommand{\hn}{\hat h_{\mathrm{n}}}

\newcommand{\No}{N_{\mathrm{o}}}

\newcommand{\tket}[1]{\ket{\tilde{#1}}}

\newcommand{\tmatel}[3]{\braket{\tilde #1 | #2 | \tilde #3}}

\begin{document}

\preprint{APS/123-QED}

\title{Time-dependent renormalized-natural-orbital theory applied to laser-driven H$_2^+$}

\author{A. Hanusch}
\author{J. Rapp}%
\author{M. Brics}%
\author{D. Bauer}%
 \email{Corresponding author: dieter.bauer@uni-rostock.de}
\affiliation{%
 Institut f\"ur Physik, Universit\"at Rostock, 18051 Rostock, Germany
}%

\date{\today}

\begin{abstract}
Recently introduced time-dependent renormalized-natural orbital theory (TDRNOT) is extended towards
a multi-component approach in order to describe H$_2^+$ beyond
the Born-Oppenheimer approximation.
Two kinds of natural orbitals, describing the electronic and the nuclear degrees of
freedom are introduced, and the exact equations of motion for them are derived.
The theory is benchmarked by comparing numerically exact results of the time-dependent Schr\"odinger equation
for a H$_2^+$ model system with the corresponding TDRNOT predictions. Ground state properties,
linear response spectra, fragmentation, and  high-order harmonic generation are investigated.
\end{abstract}

\maketitle


\section{\label{sec:intro}Introduction}
Simulating laser-driven $N$-particle systems  truly {\em ab initio}, i.e., by solving the time-dependent Schr\"odinger equation
(TDSE), is only possible for very small $N$. As more and more experiments are performed in the intense-laser, ultra-short pulse regime \cite{Schultz_14, Yamanouchi_13}, efficient time-dependent many-body methods, applicable beyond linear response, are needed.
A widely used approach is time-dependent density functional theory (TDDFT) \cite{Ullrich_11,Marques_TDDFT_06,Burke_JCP_05},
in which the single-particle density $n(\vec r, t)$ is used as the basic variable. This quantity is, in 
principle, sufficient to calculate every observable of a time-dependent quantum system \cite{Runge_PRL_84, Ullrich_11}.
However, while the scaling of the computational effort is favorable for TDDFT, a generally unknown exchange-correlation (XC)
functional is involved that needs to be approximated. Especially the often used adiabatic XC functionals often miss correlation effects \cite{Wilken_PRL_06, Ruggenthaler_PRL_09, Helbig_ChemPhys_11}.
Additionally, not all observables are known as functionals of $n(\vec r, t)$ (an example being correlated photoelectron spectra \cite{Wilken_PRA_07}), meaning that even if the exact single-particle density $n(\vec r, t)$ was reproduced by TDDFT, the interesting observables measured in nowadays intense-laser matter experiments could not be reproduced. Other approaches, e.g., 
multi-configurational time-dependent Hartree-Fock (MCTDHF) \cite{Yeager_CPL_79, Zanghellini_JPhysB_04} 
or time-dependent configuration interaction (TDCI) \cite{Krause_JCP_05, Rohringer_PRA_06, Krause_JCP_07, Greenman_PRA_10, Karamatskou_PRA_14}
do not suffer from these difficulties, however, at a price of much higher computational cost.

When applying many-body methods to molecular systems, the Born-Oppenheimer (BO) approximation is often
employed, or the nuclei are even treated classically.
However, for an accurate description of molecules in, e.g.,  strong laser fields, the nuclei should
be treated fully quantum mechanically beyond BO.
Especially in the case of fragmentation of molecules in intense laser fields the adiabatic 
BO approximation may break down as electronic and nuclear energy scales are not well separated at 
avoided crossings or conical intersections. 
Several approaches aiming at the description of correlated electron-nuclear dynamics 
beyond the BO approximation were presented in the last few years, e.g., the exact factorization 
of the molecular wavefunction~\cite{Abedi_PRL_10, Abedi_JCP_12, Suzuki_PRA_14}, a multiconfigurational time-dependent Hartree (Fock) approach [MCTDH(F)]~\cite{Nest_CPL_09, Ulusoy_JCP_12, Jhala_PRA_10}, or
 a multicomponent extension of (TD)DFT (MC(TD)DFT)~\cite{Kreibich_PRL_01, Kreibich_PRA_08, Butriy_PRA_07},
which, besides the single-particle electron density, also takes the diagonal of the nuclear density 
matrix into account. 

In this paper, we extend the recently introduced time-dependent renormalized-natural-orbital theory 
(TDRNOT)~\cite{Brics_PRA_13, Rapp_PRA_14, Brics_PRA_14, Brics_PRA_16} towards the simplest molecular system, H$_2^+$,  taking both the electronic
and nuclear degrees of freedom fully quantum mechanically into account. We restrict ourselves to a low-dimensional H$_2^+$ model system \cite{Kulander_PRA_96, Chelkowski_PRA_98, Feuerstein_PRA_03, Butriy_PRA_07, Jhala_PRA_10, Suzuki_PRA_14,Mosert}
in order to have the TDSE benchmark results readily available. However, the TDRNOT equations derived in this work are easily generalized to the ``real,'' three-dimensional (3D)  H$_2^+$.

The basic quantities of our theory are the so-called natural orbitals (NOs), introduced by L\"owdin as the eigenfunctions 
of the one-body reduced density matrix (1-RDM) \cite{Loewdin_PhysRev_55}.  Equations of motion (EOM) for the NOs can be derived. 
However, as each NO is defined up to a phase factor only, the EOM are not unique. This ``phase freedom'' can be employed to the computational benefit and to remove seeming singularities.  
Renormalizing NOs amounts to normalizing them to their eigenvalues, which simplifies an exactly unitary propagation \cite{Rapp_PRA_14}. 
TDRNOT has been applied to a model two-electron atom and performed well in treating
phenomena where TDDFT with known and practicable XC functionals fails \cite{Rapp_PRA_14, Brics_PRA_14, Brics_PRA_16}. As the NOs are proven to form the best possible basis
for two-electron systems \cite{Giesbertz_CPL_14}, the hope is that TDRNOT provides a means to treat bigger systems in a computationally economic way as well.

The paper is structured as follows. The $\mathrm{H}_2^+$ model system and the basic properties of the
reduced density matrices and NOs of a two-component system are introduced in Sec.~\ref{sec:Theory}.
The EOM for the NOs are presented in Sec.~\ref{sec:EOM}. In Sec.~\ref{sec:Results} we 
benchmark TDRNOT by first calculating ground state properties and linear response 
spectra. Second, the interaction with intense laser pulses is simulated, with the focus on 
the fragmentation dynamics and high-order harmonic generation (HHG). Finally, in Sec.~\ref{sec:Summary}
we give a conclusion.

Atomic units (a.u.) are used throughout unless noted otherwise.

\section{Natural-orbital theory\\ for a two-component system}\label{sec:Theory}

\subsection{Model system}
We apply TDRNOT to the widely used one-dimensional $\mathrm{H}_2^+$ model system 
\cite{Kulander_PRA_96, Chelkowski_PRA_98, Feuerstein_PRA_03, Butriy_PRA_07, Jhala_PRA_10, Suzuki_PRA_14,Mosert}.
This collinear model utilizes  the fact that the ionization and dissociation dynamics of H$_2^+$  is predominantly constrained to 
the polarization direction when interacting with
a strong, linearly polarized laser field. The reduced dimensionality permits the exact numerical solution of the TDSE at relatively low computational cost, and thus efficient benchmarking of TDRNOT.

The Hamiltonian of the H$_2^+$ model system (in dipole approximation and length gauge) reads
\begin{eqnarray}\label{eq:Hamiltonian}
 \hat H(x,R,t) &=& \he + \hn + V_{\mathrm{en}}(x,R) ,
\end{eqnarray}
where 
\begin{eqnarray}\label{eq:Hamiltonian1}
  \he(x,t) &=& -\frac{1}{2 \,\mu_{\mathrm{e}}}\, \partial_x^2 + q_e \, x \, E(t) \\
 \hn(R)  & = & -\frac{1}{2 \, \mu_{\mathrm{n}}}\, \partial_R^2 + V_{\mathrm{nn}}(R).\label{eq:Hamiltonian2}
\end{eqnarray}
$x$ and $R$ denote the electron coordinate and the internuclear distance, respectively.
We introduce $\he$ and $\hn$ as the single-particle Hamiltonians for the electronic and nuclear 
degree of freedom, respectively.
Furthermore, $\mu_{\mathrm{e}} = {2\,M}/{(2\,M+1)}$ (with the proton mass $M \simeq 1836$) and 
$\mu_{\mathrm{n}} = {M}/{2}$ denote the reduced masses of the electron 
and the nuclei, respectively, and $q_\mathrm{e} = {(2\,M+2)}/{(2\,M+1)}$
is the reduced charge.

The interaction potentials are modeled by soft-core potentials in order to eliminate 
the singularities:
\begin{equation}
 V_\mathrm{en}(x,R) = -\frac{1}{\sqrt{(x-\frac{R}{2})^2 + \varepsilon^2_{\mathrm{en}}}} - \frac{1}{\sqrt{(x+\frac{R}{2})^2 + \varepsilon^2_{\mathrm{en}}}} ,
\end{equation}
\begin{equation}
 V_\mathrm{nn}(R) = \frac{1}{\sqrt{R^2 + \varepsilon^2_{\mathrm{nn}}}} .
\end{equation}
The softening parameters are set to $\varepsilon^2_{\mathrm{en}} = 1$ and 
$\varepsilon^2_{\mathrm{nn}} = 0.03$.

To describe the model system in terms of NOs it is useful to expand the wavefunction in 
orthonormal single-particle wavefunctions describing the electronic and nuclear degree of freedom.
The Schmidt decomposition \cite{Pathak_13} ensures that only a single summation is necessary 
for this expansion, 
\begin{equation}\label{eq:Expansion}
 \Psi(x,R,t) = \sum_{k} c_{k}(t) \, \varphi_k(x,t) \, \eta_k(R,t) .
\end{equation}

\subsection{Density matrices and natural orbitals}
Let us start from the pure density matrix
\begin{equation}\label{eq:2-DM}
 \hat \gamma_{1,1}(t) = \ket{\Psi(t)}\bra{\Psi(t)}.
\end{equation}
Unlike in the two-electron case \cite{Rapp_PRA_14} the pure two-body density matrix (2-DM) is a multicomponent object in the case of H$_2^+$. 
Due to the two distinguishable degrees of freedom, different 1-RDMs are obtained, 
depending on which degree of freedom is traced out,
\begin{eqnarray}
 \hat \gamma_{1,0}(t) &=& \Tr_\mathrm{n} \hat \gamma_{1,1}(t), \label{eq:1-RDM_el} \\
 \hat \gamma_{0,1}(t) &=& \Tr_\mathrm{e} \hat \gamma_{1,1}(t). \label{eq:1-RDM_nuc}
\end{eqnarray}
As the NOs and occupation numbers (ONs) are defined as the eigenstates and eigenvalues of the 1-RDM, respectively,
two different kinds of orbitals are expected:
\begin{eqnarray}
\hat \gamma_{1,0}(t) \ket{k(t)} &= n_k(t) \ket{k(t)} \label{eq:Def_NOs_el}\\
\hat \gamma_{0,1}(t) \ket{K(t)} &= N_K(t) \ket{K(t)} \label{eq:Def_NOs_nuc}\!.
\end{eqnarray}
Throughout this paper, we will use lower-case letters for electronic NOs and upper-case for nuclear NOs.

Inserting Eq.~(\ref{eq:Expansion}) into Eqs.~(\ref{eq:2-DM})--(\ref{eq:Def_NOs_nuc})
leads to the conclusion that the single-particle wavefunctions in  Eq.~(\ref{eq:Expansion})
are the electronic and nuclear NOs, 
\begin{equation} \varphi_k(x,t) = \braket{x|k(t)}, \qquad \eta_k(R,t) = \braket{R|K(t)}, \end{equation} 
respectively. The expansion coefficients in 
Eq.~(\ref{eq:Expansion}) can be expressed in terms of the ONs,
\begin{equation}
 c_k(t) = \sqrt{n_k(t)} \, \ee^{\ii \phi_k(t)},
\end{equation}
i.e., they are defined up to a phase factor. Additionally, one finds the constraint 
\begin{equation}\label{eq:Condition}
 n_k(t) = N_K(t).
\end{equation}
Hence, ONs of each {\em pair} of electronic and nuclear NOs have to be equal at any time, despite their distinguishability!

For a numerical propagation it is beneficial to introduce renormalized natural orbitals (RNOs)
\begin{equation}
 \ket{\tilde k(t)} = \sqrt{n_k(t)} \ket{k(t)},\quad \ket{\tilde K(t)} = \sqrt{N_K(t)} \ket{K(t)}
\end{equation}
in order to unify the coupled equations of motion for the ONs and NOs 
and thus propagate only one combined quantity.
In terms of RNOs
\begin{equation}
 \hat \gamma_{1,0}(t) =\sum_k \ket{\tilde k(t)}\bra{\tilde k(t)}.
\end{equation}
In the same way $\hat \gamma_{0,1}(t)$ can be expanded in nuclear RNOs.
The multi-component 2-DM $\hat \gamma_{1,1}$ expanded in RNOs reads
\begin{equation}
 \hat \gamma_{1,1}(t) = \sum_{iJkL} \tilde \gamma_{iJkL}(t) \ket{\tilde i(t), \tilde J(t)} \bra{\tilde k(t), \tilde L(t)} \!.
\end{equation} 

The expansion coefficients $\tilde \gamma_{iJkL}(t)$ are exactly known in the case of a
two-particle system like helium \cite{Rapp_PRA_14}. But also for any other systems with two degrees of freedom
\begin{equation}
 \tilde \gamma_{iJkL}(t) = \frac{1}{\sqrt{n_i(t)\, n_k(t)}} \, \delta_{i,J} \, \delta_{k,L}
\end{equation}
holds.

By definitions  \eqref{eq:Def_NOs_el},  \eqref{eq:Def_NOs_nuc} the NOs are determined only up to an orbital-dependent factor. Assuming the NOs to be normalized  (e.g., to unity) there remains still the freedom to choose an orbital-dependent phase factor. Such a choice, however, will affect the phase factors $\ee^{\ii \phi_k(t)}$ in the expansion \eqref{eq:Expansion} of $\Psi(x,R,t)$.
The phase freedom of the NOs thus allows for a phase transformation leading to tunable, constant phases 
(for more details see Ref.~\cite{Rapp_PRA_14}), and all time-dependencies are then incorporated
in the so-called ``phase-including natural orbitals'' (PINOs) 
\cite{Diss_Giesbertz_10, Giesbertz_JCP_12, vanMeer_JCP_13}. Moreover, as already noted in
\cite{Rapp_PRA_14}, even after shifting all time-dependencies from the phase factor to the NOs there is still the freedom to distribute this phase arbitrarily between each pair of orbitals
in the product $\varphi_k(x,t)\,\eta_k(R,t)$.

The time-evolution of the electronic NOs can be formally expanded as
\begin{equation}
 \ii \partial_t \ket{k(t)} = \sum_m \alpha_{km}(t) \ket{m(t)}\!
\end{equation}
(analogously for the nuclear NOs). Different phase choices translate to different diagonal elements $\alpha_{kk}(t)$ and $\alpha_{KK}(t)$.

\section{Equations of motion}\label{sec:EOM}
Starting from the EOM of the 2-DM and with the knowledge of the expansions of the 1-RDM 
and 2-DM in RNOs, exact equations of motion for the two types of RNOs can be derived. 
The electronic RNOs evolve (all time arguments are suppressed for the sake of brevity) according to,
\begin{equation}
  \ii \partial_t \tket{n} = \he \tket{n} +  \mathcal{A}_n \tket{n} + \sum_{k \neq n} \mathcal{B}_{nk} \tket{k} + \sum_k \mathcal{\hat C}_{nk} \tket{k} \vphantom{\sum^\No}
\end{equation}
with the coefficients
\begin{subequations}
\begin{eqnarray}
 \mathcal{A}_n\hphantom{_l} &=& \frac{\beta_n-1}{n_n} \Re \sum_{pJL} \tilde \gamma_{nJpL} \braket{\tilde p \tilde L | \Ven | \tilde n \tilde J}\!,
 \\
 \mathcal{B}_{nk} &=& \frac{1}{n_n - n_k} \sum_{pJL} \left[ \tilde \gamma_{pLnJ} \braket{\tilde k \tilde J | \Ven | \tilde p \tilde L} \right. \nonumber\\
  & & \qquad\qquad - \left. \tilde \gamma_{kJpL} \braket{\tilde p \tilde L | \Ven | \tilde n \tilde J} \right]\!,
 \\
 \hat{\mathcal{C}}_{nk} &=& \sum_{JL} \tilde \gamma_{kJnL} \braket{\tilde L | \Ven | \tilde J}\!,
\end{eqnarray}
\end{subequations}
while the EOM for the nuclear RNOs is of a similar form
\begin{equation}
     \ii \partial_t \tket{N} = \hn \tket{N} +  \mathcal{A}_N \tket{N} + \!\! \sum_{K \neq N} \!\! \mathcal{B}_{NK} \tket{K} + \sum_K \mathcal{\hat C}_{NK} \tket{K} \vphantom{\sum^\No}
\end{equation}
with 
\begin{subequations}
\begin{eqnarray}
 \mathcal{A}_N\hphantom{_K} &=& - \frac{\beta_n}{N_N} \Re \sum_{ijL} \tilde \gamma_{iNjL} \braket{\tilde j \tilde L | \Ven | \tilde i \tilde N}\!,
 \\
 \mathcal{B}_{NK} &=& \frac{1}{N_N - N_K}  \sum_{ijL} \left[ \tilde \gamma_{jLiN} \braket{\tilde i \tilde K | \Ven | \tilde j \tilde L} \right. \nonumber\\ 
  & & \qquad\qquad - \left. \tilde \gamma_{iKjL} \braket{\tilde j \tilde L | \Ven | \tilde i \tilde N} \right]\!,
 \\   
 \hat{\mathcal{C}}_{NK} &=& \sum_{ij} \tilde \gamma_{iKjN} \braket{\tilde j | \Ven | \tilde i}\!.
\end{eqnarray}
\end{subequations}
In order to fulfill the constraint given in Eq.~(\ref{eq:Condition}) at any time also 
$\dot n_i(t) = \dot N_I(t)$ has to hold.
While this condition is automatically fulfilled during real-time propagation, the distribution of the phase
between each pair of orbitals has to be chosen in a particular way during imaginary time propagation 
in order to find the true ground state of the system. 
To that end the parameters 
\begin{equation}
 \beta_n = \frac 1 2 \Re \left[ \frac{\tmatel{N}{\hn}{N} - \tmatel{n}{\he}{n}} {\strut \sum_{k,K} \frac{1}{\sqrt{n_n \, n_k}} \braket{\tilde n \tilde N | \hat V_{en} | \tilde k \tilde K} \delta_{k,K}} + 1 \right]\!
\end{equation}
during imaginary time propagation are introduced (arbitrary real $\beta_n$ can be chosen during real time propagation; we simply took $\beta_n = 1/2$).

The EOM are exact for an infinite number of RNOs. However, in a numerical implementation  it is necessary to restrict the number of orbitals to a finite value $\No$. This truncation introduces 
errors in the propagation.
We will therefore analyze the effect of the truncation by comparing to the corresponding exact results 
obtained by propagating the full many-body wavefunction according to the TDSE.
In particular, we may extract the correct, truncation-free NOs by diagonalizing the exact 1-RDMs.

\section{Results}\label{sec:Results}
In this section, we first benchmark ground state results for the H$_2^+$ model obtained with TDRNOT in imaginary time against the TDSE result. Second, as the simplest real-time propagation application, linear response spectra are calculated for different   $\No$ and compared to the reference TDSE result. Finally, we consider the interaction with a short,
intense laser pulse.

\subsection{Ground state}
The ground state energies obtained from a TDRNOT imaginary-time propagation of $\No$ orbitals per degree of freedom 
are presented in Tab.~\ref{tab:energies}, together with the exact value from the TDSE.

\begin{table}[b]
\centering
\caption{Energies and ONs of the ground state obtained from imaginary-time propagation using different $\No$. 
The exact TDSE results are presented for comparison. With increasing $\No$ the values converge to the exact results.}
\label{tab:energies}

\begin{ruledtabular}
\begin{tabular}{lccccc}
      & Total energy             & \multicolumn{4}{c}{Dominant occupation numbers}            \\
$\No$ & $E_0 \; [\mathrm{a.u.}]$ & $n_1$ & $n_2 \,/  10^{-3}$ & $n_3 \,/ 10^{-6}$ & $n_4 \,/ 10^{-8}$ \\
\colrule  \vspace{-6pt} \\
1     & $-0.774\,84 $      & $1.000\,00$ &                    &                   &           \\
2     & $-0.776\,36 $      & $0.997\,75$ & $2.255$            &                   &           \\
4     & $-0.776\,38 $      & $0.997\,70$ & $2.291$            & $8.330$           & $4.685$   \\
8     & $-0.776\,38 $      & $0.997\,70$ & $2.291$            & $8.332$           & $4.746$   \\
TDSE  & $-0.776\,38 $      & $0.997\,70$ & $2.291$            & $8.332$           & $4.746$   \\ 
\end{tabular}
\end{ruledtabular}
\end{table}

Clearly, the TDRNOT ground state energy converges to the exact value for increasing $\No$, and only
a few RNOs are needed to obtain excellent agreement. The ONs show a behavior 
expected for the ground state: The first orbital is highly occupied with an ON close to one while 
the ONs for higher orbitals decrease rapidly with increasing orbital index.  

Using only one orbital per degree of freedom ($\No = 1$) TDRNOT corresponds to an uncorrelated time-dependent Hartree (TDH)
approach \cite{Kreibich_CP_04}. The ground-state energy is already reasonably accurate. However, it is known that the TDH approach fails to describe dissociation, as the nuclear potential 
is only well approximated around the equilibrium internuclear distance \cite{Kreibich_CP_04, Butriy_PRA_07, Jhala_PRA_10}.

\begin{figure}[t]
\includegraphics[width=\columnwidth]{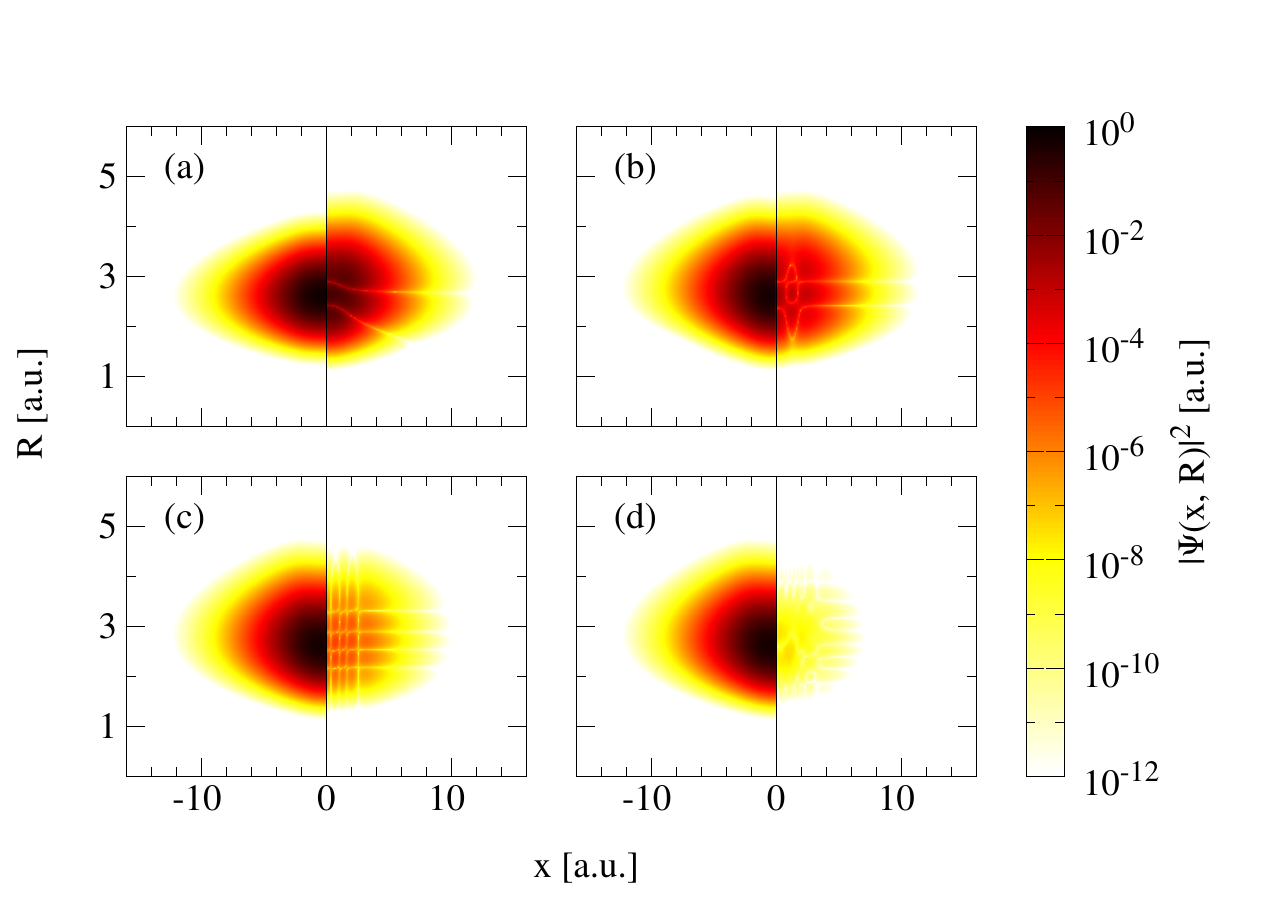}
\caption{\label{fig:GroundDensity} (color online) Plot of the correlated ground state probability density
$|\Psi(x,R)|^2$ for (a) $\No = 1$, (b) $\No = 2$, and (c) $\No = 4$ and (d) $\No = 8$ for negative values of $x$.
For $x>0$ the absolute difference to the exact probability density is plotted.}
\end{figure}

Not only the ground state energy but also the correlated ground state probability density is in excellent agreement
if enough RNOs are taken into account, as shown in Fig.~\ref{fig:GroundDensity}. 
A grid-like structure is apparent in the differences between the TDRNOT ground state probability densities and the exact  TDSE  density. This structure is related to the location of the nodal lines of the most significant
RNO not included in the TDRNOT calculation.

\subsection{Linear response spectrum}
\begin{figure}[t]
\includegraphics[width=\columnwidth]{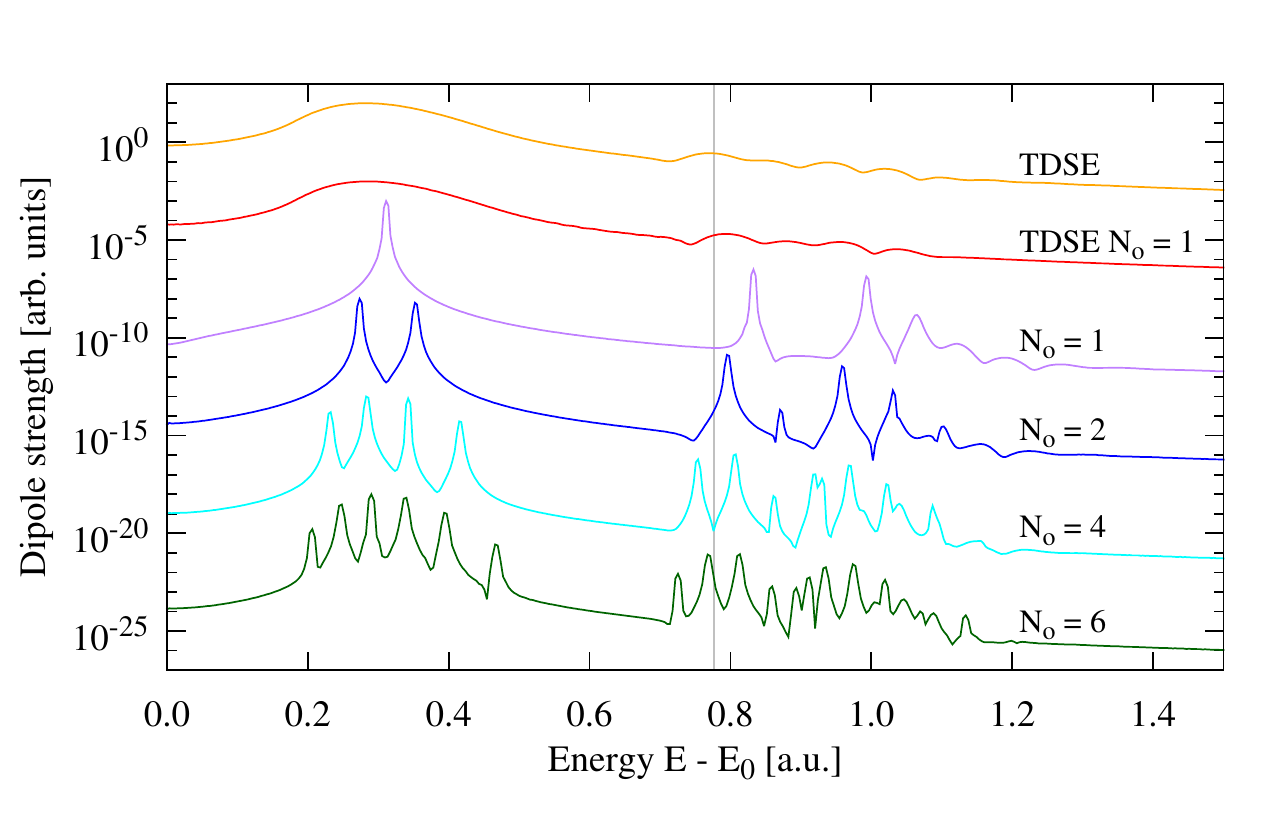}
\caption{\label{fig:LinearResponse} (color online) Linear response spectra obtained from TDRNOT calculations with
different numbers of RNOs $\No$. For comparison also the exact TDSE result is plotted.}
\end{figure}

In order
to obtain linear response  spectra, the initial ground state RNOs are propagated in real time for $t_\mathrm{max} = 2000$
after a kick with a small electric field ($E = 0.0001$). An imaginary potential is enabled to 
prevent reflection of the density at the boundaries of the grid. Fourier transforming the time-dependent 
dipole expectation value $d(t)$,
\begin{equation}
 d(t) = -\braket{\Psi(t) | q_\mathrm{e} \hat x | \Psi(t) } = -\sum_n q_\mathrm{e} \braket{\tilde n(t) | \hat x | \tilde n(t)} \!,
\end{equation}
leads to a spectrum which exhibits peaks at energy differences $E-E_0$ of 
dipole-allowed transitions. The resulting spectra calculated from TDRNOT propagations with different $\No$
as well as the reference  spectrum from a TDSE calculation are depicted in Fig.~\ref{fig:LinearResponse}.

A severe difference between the exact and the TDRNOT result is apparent.
As the electronic first excited state (in the BO-picture) is dissociative, a broad continuous 
feature is visible in the exact spectrum. This is also the case for other electronic transitions. 
In contrast to HD$^+$ \cite{Butriy_PRA_07}, vibrational excitations have vanishing dipole oscillator strengths. Hence, no excitations at low energies are visible.
The results from the TDRNOT calculations show a different behavior: Instead of a continuum
discrete peaks are visible. The number of peaks increases with the number of RNOs used in the
calculation. In contrast to the helium model atom---where including more RNOs leads to the appearance 
of peaks describing series of doubly excited states \cite{Rapp_PRA_14}---in the molecular case 
several of the emerging discrete peaks can be assigned to the same electronic transition.
The increasing number of discrete transitions should finally result in a continuous spectrum 
if enough orbitals are taken into account.
For the TDH case $\No = 1$ this behavior has already
been observed \cite{Kreibich_CP_04, Butriy_PRA_07, Jhala_PRA_10}. Using the Hartree approximation, 
only one sharp peak---corresponding to a transition to a bound state---appears in the spectrum 
for the first electronic transition. The reason for this erroneous behavior is the wrong shape 
of the nuclear potential in this case (see e.g., Refs.~\cite{Kreibich_CP_04, Butriy_PRA_07}).

As stated before, the restriction to a finite number of RNOs introduces a truncation error. 
Truncation-error-free reference results for a given $\No$ can be obtained by diagonalization of the exact 1-RDM (from the TDSE).  The resulting spectrum from only one truncation-error-free NO (labeled with TDSE $\No=1$) is also shown in Fig.~\ref{fig:LinearResponse}. It almost completely coincides with the full exact result. One thus can conclude that almost all important information 
is already included in the first dominant RNO.
However, due to the coupling between RNOs in the TDRNOT EOM all other RNOs are important during the propagation.

\subsection{H$_2^+$ in intense laser fields}

\begin{figure*}
\includegraphics[width=\textwidth]{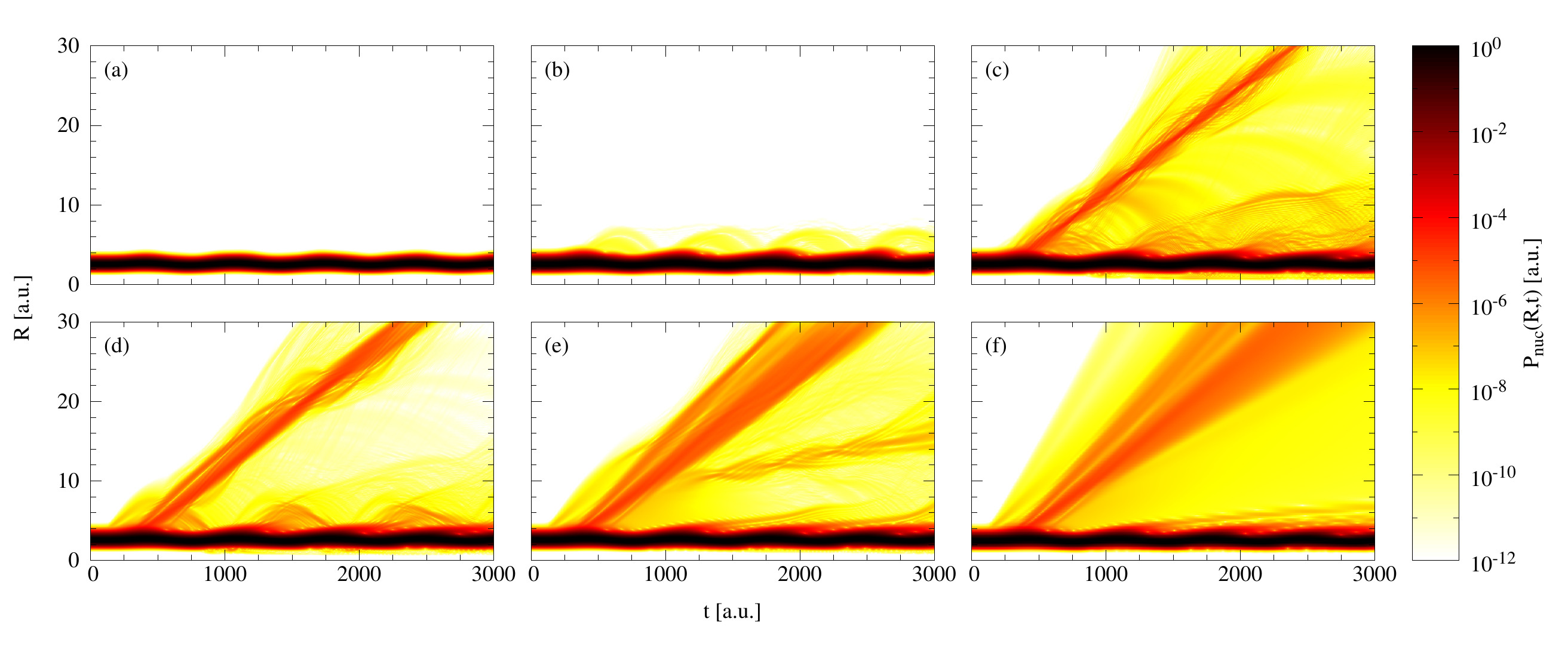}
\caption{\label{fig:NucDensity} (color online) Time-dependent nuclear probability density 
upon the interaction with a $800$-nm four-cycle pulse with  $I_0=10^{14}\, \watt\per\centi\meter^2$.
Again different numbers of orbitals were used: (a) $\No = 1$, (b) $\No = 2$, (c) $\No = 4$,
(d) $\No = 8$, and (e) $\No = 10$. With more RNOs included, the agreement with the exact 
result of the TDSE, given in panel (f), is considerably improved.}
\end{figure*}

Many different processes influence the fragmentation dynamics 
of molecules subjected to intense laser fields, e.g., bond softening \cite{Bucksbaum_PRL_90}, above-threshold dissociation (ATD) \cite{Giusti-Suzor_PRL_90}, bond hardening or vibrational trapping \cite{Giusti-Suzor_PRL_92}, 
 charge-resonance-enhanced ionization \cite{Zuo_PRA_95}, and the  ``retroaction'' due to the long-range Coulomb potential \cite{waitz}.
We want to further benchmark TDRNOT by investigating its ability to describe non-perturbative 
phenomena far from equilibrium. As the theory is aiming to describe strong-field laser-matter interaction, we study the fragmentation of H$_2^+$ upon the 
interaction with a short, intense laser pulse. Furthermore, HHG spectra are calculated.

\subsubsection{Dissociation and ionization}
An infrared $800$-nm four-cycle pulse with a $\sin^2$-envelope and a peak intensity 
of $I_0=10^{14}\, \watt\per\centi\meter^2$ was applied to the H$_2^+$ model system.
Upon the interaction with an intense laser pulse, fragmentation can occur due to dissociation or
dissociative ionization (DI). In the latter case the removal of the electron leads to Coulomb explosion
as the nuclei fly apart due to their Coulomb repulsion.
In order to judge whether the different fragmentation processes can be reproduced with TDRNOT, 
we analyze the time-dependent nuclear probability density,
\begin{equation}\label{eq:nuc_density}
 P_{\mathrm{nuc}}(R,t) = \int \mathrm{d}x \, |\Psi(x,R,t)|^2 = \sum_k | \tilde \eta_k(R,t)|^2 .
\end{equation}
Figure~\ref{fig:NucDensity} shows the logarithmically scaled, time-dependent nuclear probability density 
$P_\mathrm{nuc}(R,t)$ resulting from TDRNOT calculations. The TDSE reference result is included for comparison 
in  Fig.~\ref{fig:NucDensity}f.
In the latter figure a many fold jet-like structure becomes apparent, which can be attributed to dissociation.
Due to ATD---the absorption of more photons than needed---dissociation channels with different kinetic energies 
of the fragments appear.
In the TDH  case  $\No=1$, however, the time-dependent nuclear probability density shows no indication of 
dissociation at all (Fig.~\ref{fig:NucDensity}a). This erroneous behavior is due to the wrong shape of the 
effective nuclear potential again (see Fig.~1 in Ref.~\cite{Kreibich_CP_04}). Vibrations around the equilibrium 
internuclear distance are already reproduced though. A TDRNOT calculation with $\No = 2$ does not lead to 
a much improved result. 
However, 4 RNOs are sufficient for reproducing dissociation, as the most prominent jet is clearly visible, 
although the broadening is not yet in good agreement with $P_\mathrm{nuc}(R,t)$ obtained from the TDSE. 
As expected, including more orbitals leads to a better agreement with the exact result. A second jet 
corresponding to dissociation upon the absorption of a different number of photons is already clearly visible 
in the $\No = 8$ density, and with two more orbitals the broadening improves. 
However, an erroneous  structure emerges at intermediate internuclear distances $10<R<20$, which vanishes 
with even more RNOs (not shown).

\begin{figure}[b]
\includegraphics[width=\columnwidth]{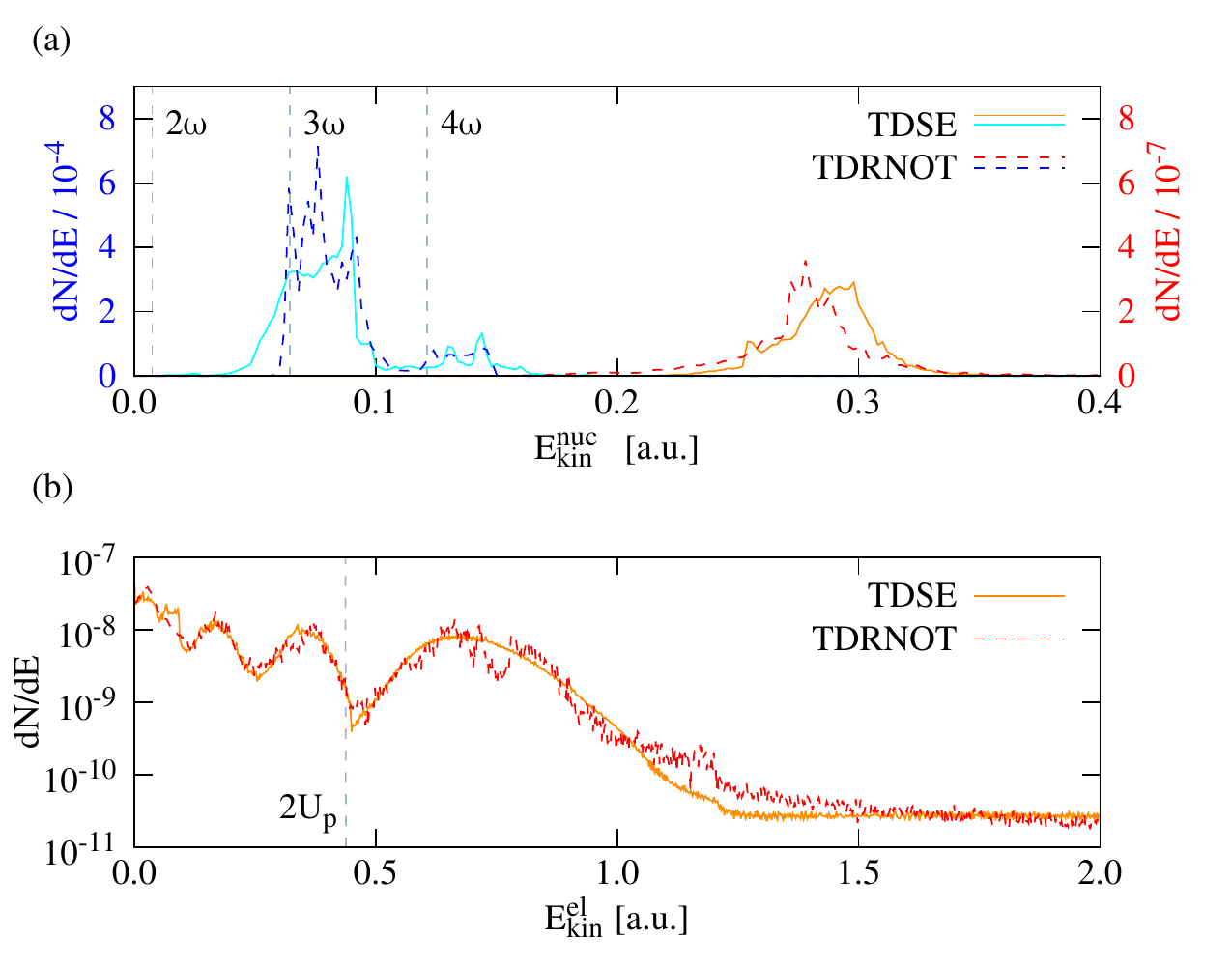}
\caption{\label{fig:EnergySpectra} (color online) Energy spectra for nuclei and photoelectrons, calculated using the (extended) virtual detector method.
(a) Kinetic-energy spectra of the nuclei for  dissociation (blue) and DI (red).
 The vertical grey lines denote $2\omega$, $3\omega$, and $4\omega$ absorption from the vibrational  ground state.
(b) TDRNOT photoelectron spectrum for 10 RNOs per degree of freedom (red, dashed) 
  compared to the exact result from the TDSE (orange, solid).}
\end{figure}

The kinetic energy release (KER) in the nuclear fragments for dissociation and DI can be calculated from the RNOs 
by means of the virtual detector method \cite{Feuerstein_JPB_2003, Feuerstein_PRA_03}.
To that end we reconstruct the wavefunction from the RNOs and then follow Ref.~\cite{Feuerstein_PRA_03}. 
The resulting KER spectra obtained with 10 RNOs per degree of freedom are compared with the corresponding TDSE 
benchmark results in Fig.~\ref{fig:EnergySpectra}a. 

Regarding dissociation, multiple peaks at energies $E_{\mathrm{kin}}<0.2$ are observed. 
The most distinct peaks are separated by roughly the photon energy and can be assigned to 
three and four-photon ATD, respectively. These processes were found to be dominant also for
longer pulses of the same wavelength and intensity~\cite{Peng_JPhysB_05}.
The expected positions of the peaks (using the BO-approximation and assuming the vibrational ground state) 
in the spectrum can be calculated using a simple energy conservation formula \cite{Chelkowski_PRA_98}. 
These positions are depicted as vertical gray lines in Fig.~\ref{fig:EnergySpectra}a.
The spectrum obtained from the TDRNOT calculation has a structure similar to the exact one---heights and positions
of the peaks coincide approximately with the exact results. 
However, in the TDRNOT spectrum several discrete peaks are visible for the three-photon dissociation
instead of the broad, continuous energy distribution in the exact spectrum.
Moreover, there are discrepancies for lower energies, and the two-photon dissociation is missing completely.

\begin{figure*}
\includegraphics[width=\textwidth]{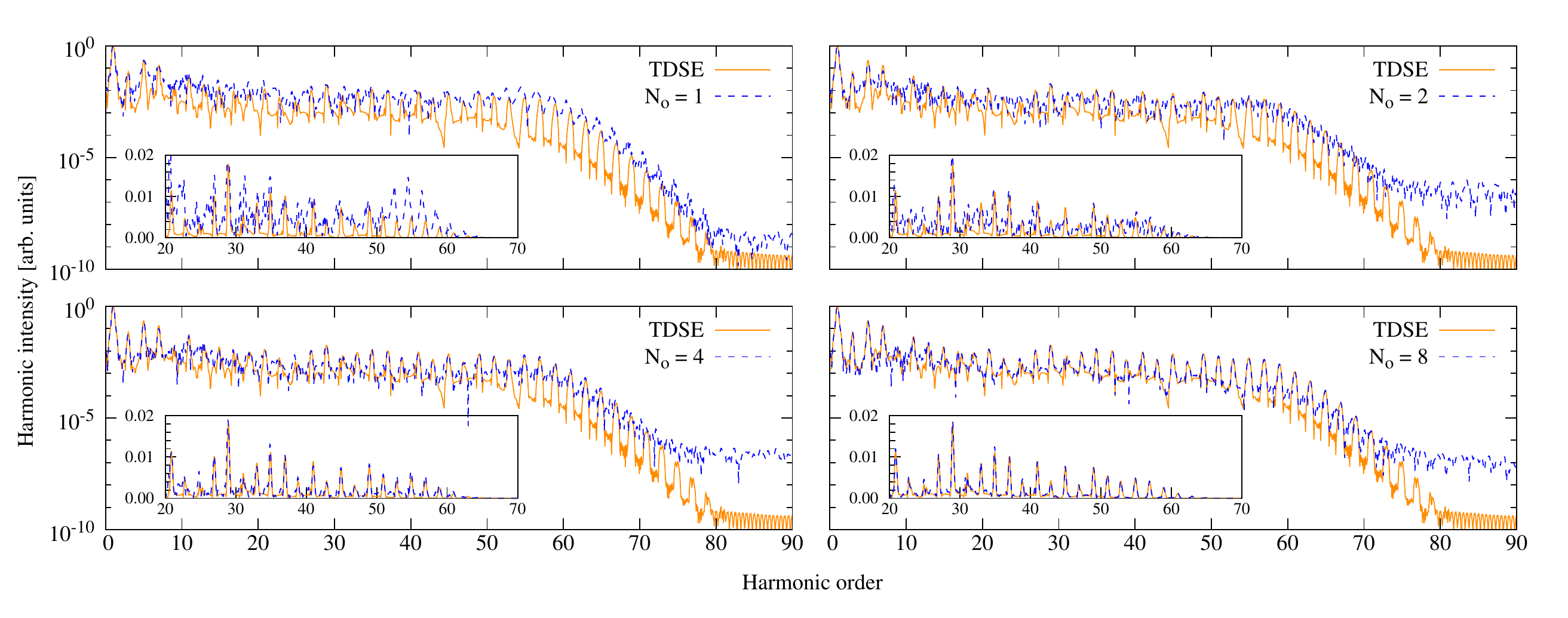}
\caption{\label{fig:HHG} (color online) HHG spectra calculated with TDRNOT using 1, 2, 4, and 8 RNOs per degree
of freedom compared to the exact spectrum obtained from the TDSE. The insets show a section of each spectrum
plotted on a linear scale.}
\end{figure*}

The KER spectrum in the case of DI is, as the Coulomb energy is released, centered around higher 
energies $E_\mathrm{kin}>0.2$. Note the different scaling of the ordinate as the ionization yield is several
orders of magnitude below the dissociation yield.
There are slight deviations of TDRNOT from the exact result---the spectrum obtained from TDRNOT is shifted 
towards lower energies---but the general structure of the spectrum is reproduced.

Furthermore, in the case of DI, we calculate electronic kinetic-energy spectra using the extended virtual 
detector method \cite{Wang_PRL_13}. Starting from the virtual detectors, classical trajectories are calculated
in order to obtain the final momentum of the electron at the end of the laser pulse. The results are presented
in Fig.~\ref{fig:EnergySpectra}b. For both, the TDRNOT and the TDSE results, a modulation in the yield,
depending on $E_\mathrm{kin}^\mathrm{el}$ is visible. This can be attributed to the interference of quantum trajectories
starting at different ionization times, which lead to the same final momentum \cite{Milosevic_JPhysB_06, Arbo_PRA_10}.
In the case of the electronic kinetic-energy spectrum, the agreement between the results from a TDRNOT 
calculation with $\No = 10$ and the exact result is clearly better than for the KER spectra.
This shows that different minimum numbers of RNOs are required, depending on the observable to calculate.

\subsubsection{HHG spectra}
Harmonic spectra are obtained by Fourier transforming the time-dependent dipole acceleration $\ddot{d}(t)$ 
\cite{Burnett_PRA_92}, which is given by
\begin{equation}
 \mu_\mathrm{e} \, \ddot{d}(t) = \braket{ \Psi (t) | -\nabla \Ven + q_\mathrm{e} \, E(t) | \Psi (t) } .
\end{equation}
An $800$-nm 10-cycle pulse with $\sin^2$-shaped 
on- and off-ramping over two cycles was employed. The peak intensity of the laser pulse was 
$I_0 = 3.0 \times 10^{14} \, \watt\per\centi\meter^2$.
In Fig.~\ref{fig:HHG}, TDRNOT HHG spectra, calculated using 1 to 8 RNOs per degree of 
freedom, are compared to the exact TDSE spectrum. In the inset, 
a part of the spectrum is plotted on a linear scale. 

With only 1 RNO 
the position of the cut-off is already in good agreement with the exact result. 
However, the shape of individual peaks, especially at high harmonic order, is completely wrong. 
The TDRNOT calculation with $\No = 2$ exhibits erroneous
peaks in addition to the peaks at the odd harmonics, especially pronounced in the region 
beyond the cut-off.
When adding more RNOs the quantitative agreement improves, and the wrong peaks vanish. A similar improvement with increasing number of 
single-particle functions has been reported for calculations using a MCTDH approach \cite{Jhala_PRA_10}. 
For $\No = 8$ the height and the shape of the peaks are well reproduced up to the 60th 
harmonic order. At very high harmonic orders  some deviations in the spectra are still 
visible, and the noise level of the TDRNOT results is two orders of magnitude higher than for the TDSE. A similar behavior was observed for HHG in a model He atom \cite{Brics_PRA_16}.
On a linear scale, 
as often used in experiments, the agreement is excellent and clearly improves with increasing $\No$ (see insets in Fig.~\ref{fig:HHG}).

\section{Conclusion}\label{sec:Summary}
We have investigated the performance of time-dependent renormalized-natural-orbital theory (TDRNOT) when applied to the simplest multi-component system exhibiting   electron-nuclear correlation, i.e., H$_2^+$.
 Different types of renormalized natural orbitals (RNOs), describing the electronic and the nuclear component,   were introduced, 
 and their coupled EOM derived.
As in the case of helium investigated earlier 
no approximations concerning the expansion of the time-dependent two-body density matrix need to be made.

In order to benchmark the theory the ground state of a one-dimensional H$_2^+$  model system and linear response spectra were calculated using TDRNOT. 
While an excellent agreement with the exact ground state energy was achieved with very few orbitals, the linear response spectra were plagued by multiple sharp peaks that only for very many orbitals would reproduce the correct, broad structure  caused by bound-continuum transitions. This unpleasant feature is caused by the restriction to a finite 
number of orbitals, which introduces a truncation error. Future work will be devoted to improve on that aspect of TDRNOT.

Finally,  TDRNOT was applied to H$_2^+$ interacting with a short, 
intense laser pulse. The time evolution of the nuclear probability density 
was studied, and features indicating different 
fragmentation processes were identified. It was found that TDRNOT is able to reproduce dissociation and Coulomb 
explosion and the corresponding kinetic-energy-release spectra if enough RNOs are taken into account. The same applies to high-harmonics spectra where 8 RNOs were found yield very good agreement  with the benchmark result
from the time-dependent Schr\"odinger equation.

\medskip

\begin{acknowledgments}
This work was supported by the Collaborative Research Center SFB 652 of the German Science Foundation (DFG).
\end{acknowledgments}

\bibliography{hanusch_literature}

\end{document}